\documentclass[sigconf]{acmart}

\usepackage{algorithm}
\usepackage{algorithmic}
\usepackage{balance}
\usepackage{multirow}
\usepackage{amsmath}
\usepackage{mathtools}
\usepackage{amsthm}

\usepackage{booktabs}
\usepackage{xspace}
\newcommand{\ours}[0]{RQFedRec\xspace}

\AtBeginDocument{%
  }

\copyrightyear{2026}
\acmYear{2026}
\setcopyright{cc}
\setcctype{by}
\acmConference[CIKM '26]{Proceedings of the 35th ACM International Conference on Information and Knowledge Management}{November 07--11, 2026}{Rome, Italy}
\acmBooktitle{Proceedings of the 35th ACM International Conference on Information and Knowledge Management (CIKM '26), November 07--11, 2026, Rome, Italy}
\acmDOI{10.1145/3799682.3841023}
\acmISBN{979-8-4007-2539-5/2026/11}

\begin{document}

\title{Dynamic Feature-Embedding Communication via Codebook Distillation for Federated Recommendation}

\author{Mingzhe Han}
\orcid{0000-0002-4911-6093}
\affiliation{
  \institution{Fudan University}
  \city{Shanghai}
  \country{China}
}
\email{mzhan22@m.fudan.edu.cn}

\author{Jiahao Liu}
\orcid{0000-0002-5654-5902}
\affiliation{%
  \institution{Fudan University}
  \city{Shanghai}
  \country{China}
}
\email{jiahaoliu21@m.fudan.edu.cn}

\author{Dongsheng Li}
\orcid{0000-0003-3103-8442}
\affiliation{
  \institution{Microsoft Research Asia}
  \city{Shanghai}
  \country{China}
}
\email{dongshengli@fudan.edu.cn}

\author{Jiankui Zhou}
\orcid{0009-0003-6986-1300}
\affiliation{%
  \institution{Fudan University}
  \city{Shanghai}
  \country{China}
}
\email{jkzhou24@m.fudan.edu.cn}

\author{Yaqiong Li}
\orcid{0000-0001-9253-5878}
\affiliation{%
  \institution{Fudan University}
  \city{Shanghai}
  \country{China}
}
\email{liyq22@m.fudan.edu.cn}

\author{Hansu Gu}
\orcid{0000-0002-1426-3210}
\affiliation{
  \institution{Independent}
  \city{Seattle}
  \country{United States}
}
\email{hansug@acm.org}

\author{Peng Zhang}
\orcid{0000-0002-9109-4625}
\authornote{Corresponding author.}
\affiliation{
  \institution{Fudan University}
  \city{Shanghai}
  \country{China}
}
\email{zhangpeng\_@fudan.edu.cn}

\author{Ning Gu}
\orcid{0000-0002-2915-974X}
\affiliation{
  \institution{Fudan University}
  \city{Shanghai}
  \country{China}
}
\email{ninggu@fudan.edu.cn}

\author{Tun Lu}
\orcid{0000-0002-6633-4826}
\authornotemark[1]
\affiliation{
  \institution{Fudan University}
  \city{Shanghai}
  \country{China}
}
\email{lutun@fudan.edu.cn}

\renewcommand{\shortauthors}{Mingzhe Han et al.}

\begin{abstract}
Federated recommendation systems commonly protect user privacy by keeping user parameters on local devices, while exchanging item parameters for collaborative model training.
However, such item parameters usually model items independently and suffer from both efficiency and effectiveness challenges, making communication costs grow with the item space and limiting cross-item generalization and robustness to noisy feedback.
To address these limitations, we propose to model items via shared latent feature embeddings for communication.
Residual Quantization (RQ) provides a natural way to instantiate this communication by representing each item with a short sequence of discrete code IDs, i.e., Semantic IDs (SIDs).
However, directly applying centralized and static RQ-based recommendation to federated learning is non-trivial due to 1) private and biased historical interactions and 2) evolving collaborative information.
We propose \textbf{RQFedRec}, an RQ-based federated recommendation framework for dynamic feature-embedding communication.
To construct globally aligned codebooks without accessing private interactions, RQFedRec introduces an \textbf{information distillation module}.
Each client first learns item embeddings that encode local collaborative information from private interactions, and then distills such information into feature-indexed codebooks under globally shared SIDs, making sparse and biased local signals more compatible with server aggregation.
To adapt to evolving collaborative information, RQFedRec introduces a \textbf{self-refining SID update module} that dynamically refines global SID assignments from aggregated codebooks.
Extensive experiments demonstrate that RQFedRec improves recommendation performance and reduces communication costs without relying on semantic information, while further benefiting from public semantics when available.
Code is available at https://github.com/Mingzhe-Han/RQFedRec.
\end{abstract}



\maketitle

\begin{figure}[h]
\centering
\includegraphics[width=1\columnwidth]{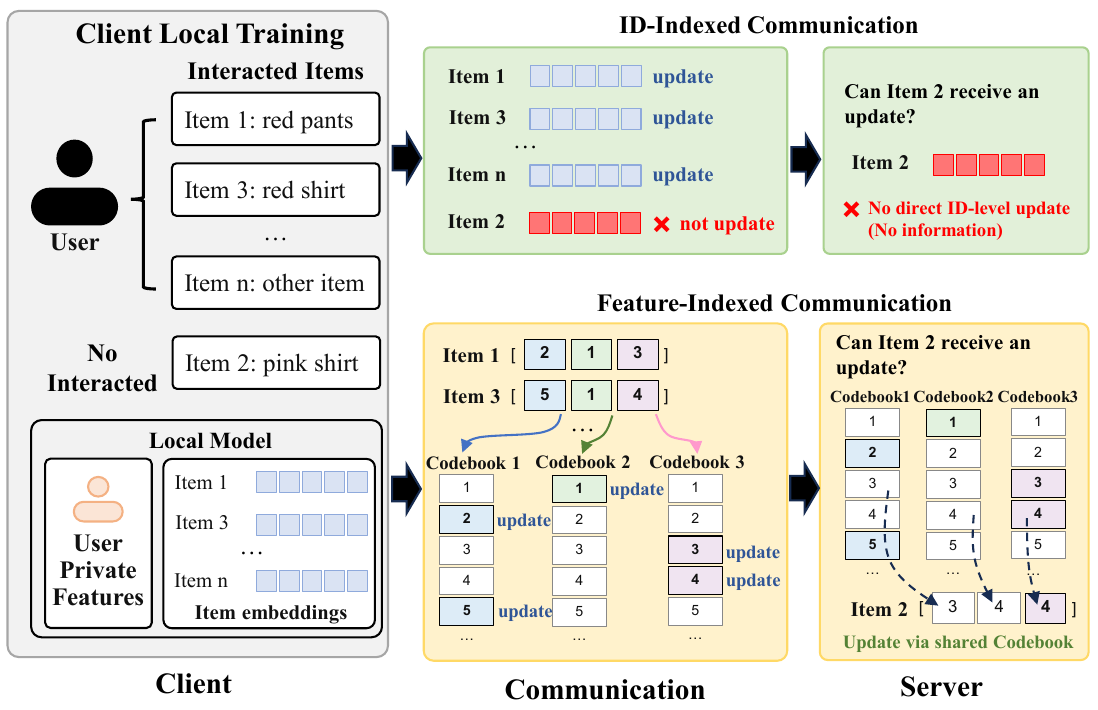}
\caption{Illustration of two communication paradigms in federated learning.} 
\label{fig:0}
\end{figure}

\section{Introduction}

Recommendation algorithms~\cite{liu2023personalized,liu2023triple} aim to deliver personalized content from historical user interaction data~\cite{zang2025mutual,zang2022survey,zhao2023task}.
However, user behaviors often contain sensitive personal information~\cite{mcsherry2009differentially,zhao2025social,gu2026inter}.
With the increasing enforcement of data protection regulations~\cite{voigt2017eu}, it becomes difficult to directly collect raw interaction data on a central server.
Federated learning~\cite{li2020federated,wen2023survey} offers a promising alternative by training models on distributed client devices and transmitting model information, thereby enabling recommendation while keeping user data local.

Existing federated recommendation algorithms~\cite{chai2020secure,zhang2024gpfedrec,han2025fedcia} commonly rely on item information for communication.
Specifically, due to the privacy risks of user preferences, user parameters are usually kept on local devices, while item parameters are communicated between clients and the server.
In recommendation algorithms, these item parameters are mainly item embeddings~\cite{zhang2023dual}, where each row is an embedding of a specific item ID~\cite{he2017neural}.
Therefore, the communicated information consists of such ID-indexed embeddings; we refer to this setting as the \textbf{ID-indexed} communication paradigm.
This paradigm is compatible with existing recommendation algorithms and easy to deploy in practical federated recommendation.

Despite its popularity, ID-indexed item communication is limited in federated recommendation in terms of both \textbf{efficiency} and \textbf{effectiveness}.
For efficiency, this paradigm transmits item information through independent item embeddings.
Thus, the communication cost increases with the number of items, making frequent transmission expensive in large-scale recommender systems~\cite{zhang2023lightfr}.
For effectiveness, such communication updates item embeddings associated with isolated item IDs, which limits cross-item generalization and robustness to noisy feedback.
As illustrated in Figure~\ref{fig:0}, if a user only interacts with Item 1 (red pants) and Item 3 (red shirt), the server only receives updates for these two item IDs.
Although Item 2 (pink shirt) may be related to the interacted items, it is not involved in this communication. Therefore, this user behavior cannot increase the model's preference estimate for Item 2 through the uploaded item updates.
Moreover, when a user accidentally clicks an item, the noisy feedback is directly reflected in its corresponding item embedding.
This item-level update can easily bias the representation of the item, especially when the item is long-tailed and has only a few interactions~\cite{liu2020long}.

These limitations motivate a \textbf{feature-indexed} communication paradigm, where item information is communicated through shared latent features from collaborative information instead of item embeddings indexed by isolated item IDs.
Recent Residual Quantization (RQ)-based methods~\cite{lee2022autoregressive,rajput2023recommender} provide a possible tool: an item can be represented by a short sequence of discrete IDs, where each ID corresponds to a latent feature and can be shared by multiple items.
Such an ID sequence is usually referred to as the Semantic ID (SID), and each position in the SID corresponds to a codebook, which stores the code embeddings (i.e., the embeddings selected by SIDs at this position).
For the efficiency limitation, the communicated parameters are determined by the size of codebooks rather than the number of item IDs, and new items can be assigned to existing codebooks without introducing a new independent item embedding.
For the effectiveness limitation, related items that the user has not interacted with can still receive updates through shared codebooks.
As illustrated in Figure~\ref{fig:0}, Item 2 can also be updated because the user interacted with Item 3, which shares part of the codebook embedding with Item 2.
Moreover, each code embedding aggregates information from multiple items, making the communication less sensitive to isolated noisy feedback.

However, RQ-based recommendation cannot be directly combined with federated learning.
Existing RQ-based methods usually construct SIDs and codebooks from item semantics or collaborative information learned from centralized historical interactions~\cite{zhou2025onerec}.
The generated SIDs are then used as static item features for downstream recommendation model training.
This centralized and static pipeline raises two challenges under federated constraints.
\textbf{First, private and biased interactions make centralized SID construction difficult.}
Constructing centralized SIDs requires user interaction data to capture collaborative information, but such data are private and cannot be directly uploaded.
Moreover, sparse and biased local interactions make it difficult to construct globally aligned SIDs and codebooks.
\textbf{Second, federated training makes static SIDs insufficient.}
Federated recommendation is an iterative process where collaborative information is continuously encoded into item embeddings.
Thus, SIDs and codebooks constructed before training may become misaligned with the updated collaborative information, making dynamic refinement necessary.

Motivated by the above observations, we propose \ours, an RQ-based federated recommendation framework for \textbf{feature-indexed} communication.
\ours replaces the upload of ID-indexed item embeddings with compact feature-indexed codebooks while keeping the local recommendation backbone (e.g., MF) unchanged.
In each communication round, clients upload local codebooks to the server for aggregation, and the aggregated codebooks are then broadcast back for the next round of local training.
Rather than simply combining RQ-based recommendation with federated learning, \ours introduces two key designs.
First, the \textbf{information distillation module} addresses the difficulty of learning globally aligned codebooks from private and biased interactions.
Instead of directly optimizing codebooks with sparse local recommendation loss, each client first learns item embeddings as flexible local representations to capture collaborative information, and then distills such information into local codebooks under shared SIDs.
This aligns locally learned collaborative information for server aggregation without uploading raw interactions.
Second, the \textbf{self-refining SID update module} addresses the limitation of static SIDs.
Rather than keeping SIDs fixed throughout training, the server periodically refines SIDs based on aggregated codebooks, allowing global SIDs to adapt to the updated collaborative information during federated training.

Our main contributions are summarized as follows:
\begin{itemize}
    \item We propose \textbf{feature-indexed} communication to address the limitations of conventional ID-indexed item communication in federated recommendation. Instead of transmitting item embeddings tied to isolated item IDs, feature-indexed communication transmits shared feature embeddings that can be reused by multiple items.
    \item We instantiate feature-indexed communication with Residual Quantization and propose \textbf{\ours}. We design an information distillation module to transfer locally learned collaborative information into shared codebooks, together with a self-refining SID update module to adapt the feature-indexed structure during federated learning.
    \item We conduct extensive experiments on multiple real-world datasets. The results show that \ours improves recommendation performance and communication efficiency without relying on semantic priors, while further benefiting from public semantics when available.
\end{itemize}

\section{Related Work}

\subsection{Federated Learning}

Federated Learning (FL)~\cite{yang2019federated} is a distributed learning framework that enables multiple clients to collaboratively train a shared model without the collection of raw data.
FedAvg~\cite{mcmahan2017communication} aggregates client updates by weighted averaging according to the data size of each client.
Building upon FedAvg, many studies focus on improving server aggregation~\cite{wang2020tackling} and stabilizing client optimization~\cite{li2020heterogeneous} under statistical heterogeneity.
In addition, some FL methods focus on client selection~\cite{lai2021oort} to improve stability and update sparsification~\cite{aji2017sparse} to reduce the communication cost.
Despite these advances, those FL algorithms primarily target supervised learning and do not explicitly address recommendation scenarios, where parameters are linearly related to the number of items and interactions are highly sensitive to noise.

\subsection{Federated Recommendation Algorithms}

Federated recommendation~\cite{sun2024survey,tan2020federated,zhang2026transfr} adapts the FL framework to recommendation algorithms.
Early work such as FedRec~\cite{lin2020fedrec} masks user interactions by sampling unrated items and using virtual ratings to reduce privacy leakage.
FedMF~\cite{chai2020secure} further protects gradients via secure aggregation such as homomorphic encryption.
Subsequent methods like FedNCF~\cite{perifanis2022federated} extend the idea to neural collaborative filtering.

To improve personalization in federated recommendation, recent studies explore personalized federated learning mechanisms.
PFedRec~\cite{zhang2023dual} introduces a dual personalization strategy to optimize both global and client models.
GPFedRec~\cite{zhang2024gpfedrec} leverages user relationship graphs to capture inter-user correlations.
FedRAP~\cite{li2024federated} enhances personalization by decomposing item embeddings into a global component learned via federation and a local additive component learned on each client.
FedCIA~\cite{han2025fedcia} emphasizes the aggregation of collaborative information rather than directly averaging item parameters.
While these approaches enable privacy-preserving training, they often upload ID-indexed item embeddings for aggregation, which scales poorly with the number of items, generalizes weakly across related items, and is sensitive to noisy feedback~\cite{gu2026llm}. 

\subsection{RQ-based Recommendation}

Recent generative recommendation studies have explored residual-quantized codebooks and Semantic IDs (SIDs) to represent items as short discrete token sequences~\cite{rajput2023recommender,tong2026rq}.
Such discrete IDs have attracted increasing attention in generative recommendation, since they reduce the item output space and make the recommendation compatible with sequence modeling architectures.
However, existing RQ-based recommendation methods~\cite{wang2024learnable} are mainly designed for centralized and static recommendation rather than federated communication.
They usually assume that historical interactions or collaborative item embeddings are available for SID construction, and the constructed SIDs remain fixed during downstream training~\cite{xiao2025unger}.
This assumption does not hold in federated recommendation, where interactions are private and distributed, and collaborative item information evolves during iterative training.
In addition, FedMM~\cite{zhang2026fedmm} encodes user and item information into a codebook for the CTR task, which is not suitable for our top-N retrieval setting and ignores dynamic collaborative information in federated training.
As a result, SIDs learned by existing methods cannot be directly used as a globally aligned and dynamically adaptive communication structure in federated recommendation.

\begin{figure*}[t]
\centering
\includegraphics[width=2.0\columnwidth]{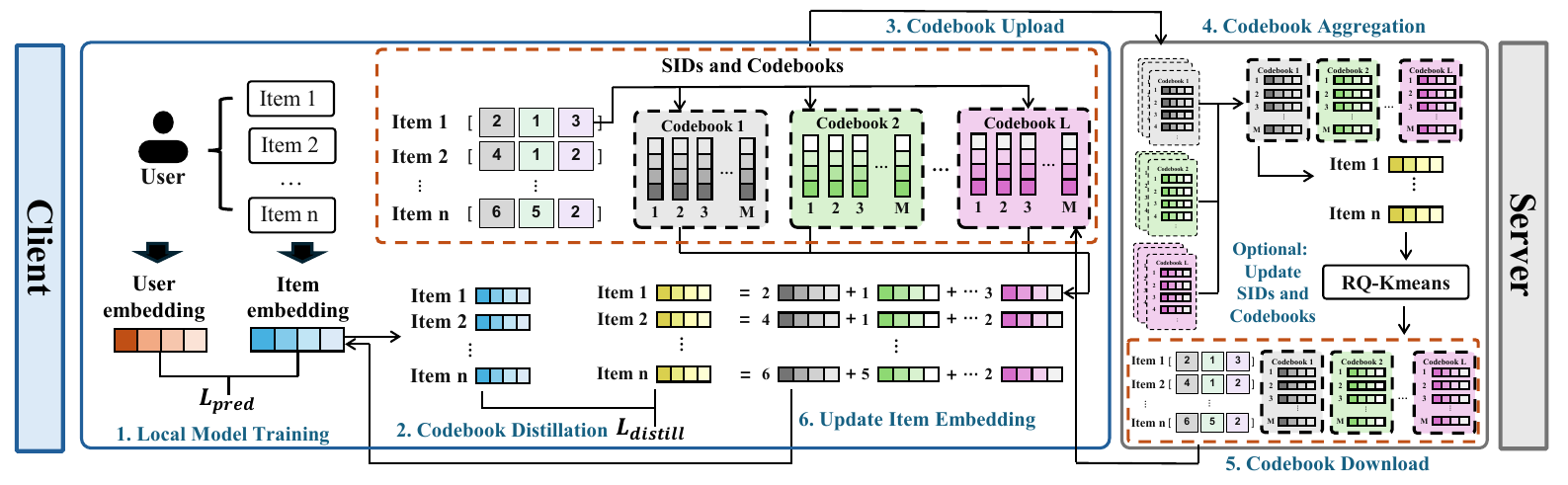}
\caption{
Overview of the \ours framework. We only illustrate a single client for simplicity.
}
\label{fig:fra}
\end{figure*}

\section{Preliminaries}
\label{sec:pre}

\subsection{Recommendation Algorithms}

Given a user set $\mathcal{U}$ and an item set $\mathcal{I}$, a recommendation algorithm aims to predict user-item interactions based on an interaction dataset $\mathcal{D}=\{(u,i,y_{ui})\}$, where $(u,i)$ is a user-item pair and
$y_{ui}\in\{0,1\}$ indicates whether user $u$ interacted with item $i$.

Most recommendation algorithms can be represented as a prediction function $R(\cdot)$ that outputs a prediction score $\hat{y}_{ui}$:
\begin{equation}
\hat{y}_{ui}=R(u,i)=P\!\left(E_u(u;\theta_u),\,E_i(i;\theta_i);\theta_p\right),
\end{equation}
where $E_u(\cdot;\theta_u)$ and $E_i(\cdot;\theta_i)$ are the user and item embedding layers with parameters $\theta_u$ and $\theta_i$, respectively, and $P(\cdot;\theta_p)$ is the prediction function, such as a dot product or MLP layer parameterized by $\theta_p$.
We denote all model parameters as $\theta=\{\theta_u,\theta_i,\theta_p\}$.

\subsection{Federated Recommendation Algorithms}

In federated recommendation, user-item interactions are stored locally to protect user privacy.
We assume there are $K$ clients or devices, indexed by $k\in\{1,\dots,K\}$, and client $k$ holds a local dataset $\mathcal{D}_k\subset \mathcal{D}$.
The goal of federated recommendation is to learn a global recommendation model without collecting user-item interactions on the server.
In conventional federated learning methods such as FedAvg~\cite{mcmahan2017communication}, each client $k$ first trains its local parameters $\theta_k$ based on its local dataset $\mathcal{D}_k$, and the server aggregates client models by weighted averaging:
\begin{equation}
\theta^{t+1}
=
\sum_{k=1}^{K}
\frac{|\mathcal{D}_k|}{\sum_{j=1}^{K}|\mathcal{D}_j|}
\theta_k^{t+1},
\end{equation}
where $\theta^{t+1}$ denotes the global parameters after aggregation in round $t$.

In federated recommendation, directly uploading all model parameters may expose private user information.
A common practice is partial parameter upload~\cite{lin2020fedrec}, where user parameters $\theta_u$ are kept on local devices, while the server only aggregates item parameters or public parameters. 
Since the uploaded item parameters are usually organized according to item IDs, we refer to this setting as ID-indexed communication.

\subsection{Residual Quantization k-means}
\label{sec:pre:rq}

Residual Quantization k-means (RQ-Kmeans)~\cite{chen2010approximate} is a multi-level vector quantization method that represents each embedding by a sequence of discrete code IDs.
Given an item embedding table $\mathbf{E}=\{\mathbf{e}_i\}_{i\in\mathcal{I}}$, where $\mathbf{e}_i\in\mathbb{R}^d$, RQ-Kmeans generates $L$ codebooks 
$\{\mathcal{B}^{(\ell)}\}_{\ell=1}^{L}$.
Each codebook
$\mathcal{B}^{(\ell)}=\{\mathbf{b}^{(\ell)}_1,\dots,\mathbf{b}^{(\ell)}_M\}\subset\mathbb{R}^d$
contains $M$ code embeddings.
Each item is encoded into a length-$L$ tuple of discrete code IDs:
\begin{equation}
\mathbf{q}_i
=
\big(q_i^{(1)},\dots,q_i^{(L)}\big),
\qquad
q_i^{(\ell)}\in\{1,\dots,M\}.
\end{equation}
We refer to the code ID tuple $\mathbf{q}_i$ as the Semantic ID (SID) of item $i$.
Given the SID $\mathbf{q}_i$, the item embedding $\mathbf{e}_i$ can be approximately reconstructed by summing the selected code embeddings:
\begin{equation}
\tilde{\mathbf{e}}_i
=
\sum_{\ell=1}^{L}
\mathbf{b}^{(\ell)}_{q_i^{(\ell)}}.
\end{equation}

RQ-Kmeans encodes each item embedding sequentially by quantizing the residual left by previous codebooks.
Specifically, it takes the original embedding $\mathbf{e}_i$ as the first-layer residual: $\mathbf{r}^{(1)}_i = \mathbf{e}_i$.
At the $\ell$-th layer, k-means is applied to the residual embeddings $\{\mathbf{r}^{(\ell)}_i\}_{i\in\mathcal{I}}$, and the cluster centers are used as the code embeddings 
$\{\mathbf{b}^{(\ell)}_1,\dots,\mathbf{b}^{(\ell)}_M\}$ 
in codebook $\mathcal{B}^{(\ell)}$.
Each item $i$ is assigned to the nearest code embedding:
\begin{equation}
q_i^{(\ell)}
=
\arg\min_{j\in\{1,\dots,M\}}
\left\|
\mathbf{r}^{(\ell)}_i-\mathbf{b}^{(\ell)}_j
\right\|_2^2.
\end{equation}
Then, the residual embedding for the next layer is updated as:
\begin{equation}
\mathbf{r}^{(\ell+1)}_i
=
\mathbf{r}^{(\ell)}_i
-
\mathbf{b}^{(\ell)}_{q_i^{(\ell)}}.
\end{equation}
After $L$ layers, RQ-Kmeans obtains a SID $\mathbf{q}_i$ for each item and a set of codebooks for reconstructing item embeddings.

\section{Methodology}
\label{sec:method}

\subsection{Overview}

In this section, we introduce \ours, an RQ-based federated recommendation framework that replaces ID-indexed item embedding communication with \textbf{feature-indexed} codebook communication.
Instead of uploading item embeddings directly, \ours represents each item by a SID sequence and communicates the shared codebooks associated with these SIDs.
Figure~\ref{fig:fra} illustrates one communication round of \ours.

Each client first trains its local recommendation model on private interactions.
After local training, the \textbf{information distillation module} distills the collaborative information captured by the learned item embeddings into local codebooks under the current SIDs.
Then, each client uploads its local codebooks to the server.
The server aggregates the uploaded codebooks to obtain updated global codebooks.
After every $\tau$ communication rounds, the \textbf{self-refining SID update module} reconstructs global item embeddings from the aggregated codebooks and updates SIDs and codebooks.
Finally, the updated codebooks and SIDs are broadcast to clients for the next round, where clients refresh their local item embeddings and continue training.

\subsection{SID and Codebook Initialization}

Before federated training, \ours needs to initialize globally shared SIDs and global codebooks for feature-indexed communication.
The global SIDs specify how each item selects feature embeddings from codebooks, while the global codebooks store the shared feature embeddings communicated between clients and the server.
After initialization, the server broadcasts the SIDs and codebooks to all clients.
In this way, all clients learn and upload codebooks under the same index space, which enables consistent aggregation of the uploaded codebooks.

Formally, let the global codebooks at round $t$ be
\begin{equation}
\mathcal{B}^{t}
=
\left\{
\mathbf{B}^{t,(\ell)} \in \mathbb{R}^{M\times d}
\right\}_{\ell=1}^{L},
\end{equation}
where $L$ is the number of RQ layers, $M$ is the number of code embeddings in each codebook, and $d$ is the embedding dimension.
Each item $i$ is assigned a length-$L$ SID:
\begin{equation}
\mathbf{q}_i^{t}
=
\big(q_i^{t,(1)},\dots,q_i^{t,(L)}\big),
\qquad
q_i^{t,(\ell)}\in\{1,\dots,M\}.
\end{equation}
Given the global codebooks and the SID of item $i$, its embedding can be reconstructed as
$
\tilde{\mathbf{v}}_i^{t}
=
\sum_{\ell=1}^{L}
\mathbf{B}^{t,(\ell)}
\big[
q_i^{t,(\ell)}
\big],
$
where $\mathbf{B}^{t,(\ell)}[q_i^{t,(\ell)}]$ denotes the selected code embedding from the $\ell$-th codebook.

The global codebooks are randomly initialized in the collaborative embedding space:
\begin{equation}
\label{eq:random_codebook_init}
\mathbf{B}^{0,(\ell)}[m]
\sim
\mathcal{N}(\mathbf{0}, \sigma_b^2\mathbf{I}),
\qquad
\ell=1,\dots,L,\quad m=1,\dots,M,
\end{equation}
For SIDs, \ours supports semantic initialization and random initialization.
When public item content features are available, we obtain an initial semantic item embedding $\mathbf{s}_i\in\mathbb{R}^{d_s}$ for each item $i$ from these public features.
The server applies RQ-Kmeans to the semantic embeddings $\{\mathbf{s}_i\}_{i\in\mathcal{I}}$ and retains the resulting discrete code assignments as the initial global SIDs $\mathbf{q}^{0}$.
Semantic SID initialization provides a useful starting structure for federated communication, especially when collaborative information is insufficient in early training.

However, in many recommendation scenarios, reliable semantic information may be unavailable or noisy. Moreover, using external semantic features may make the comparison with existing federated recommendation baselines unfair.
Therefore, \ours also supports randomly initializing SIDs without any external semantic information:
\begin{equation}
\label{eq:random_sid_init}
q_i^{0,(\ell)}
\sim
\operatorname{Uniform}\{1,\dots,M\},
\qquad
i\in\mathcal{I},\quad \ell=1,\dots,L.
\end{equation}
This setting allows \ours to be trained purely from interaction data and helps to verify that the proposed feature-indexed communication framework does not rely on external semantic information.

It is worth noting that both SID initialization strategies only provide the starting assignments for feature-indexed communication.
During federated training, the codebooks are updated by the information distillation module, and the SIDs are periodically refined by the self-refining SID update module.
Thus, the SIDs and codebooks can gradually adapt to collaborative information learned from clients regardless of the SID initialization strategy.

\subsection{Communication Pipeline}

In this section, we describe the training pipeline in a single communication round. For simplicity, we omit the round superscript $t$ when there is no ambiguity.

\subsubsection{Local Model Training}

Before local training in the current round, each client has received the global codebooks $\mathcal{B}$ and SIDs $\mathbf{q}$ broadcast by the server, and initializes its local item embeddings accordingly.
Client $k$ then trains its local recommendation model on its private dataset $\mathcal{D}_k$ by minimizing a prediction loss:
\begin{equation}
\label{equ:loss_pred}
\mathcal{L}^{(k)}_{\mathrm{pred}}(\theta_k)
=
\frac{1}{|\mathcal{D}_k|}
\sum_{(u,i,y)\in\mathcal{D}_k}
\ell_{\mathrm{pred}}\!\left(R(u,i),y\right),
\end{equation}
where $\ell_{\mathrm{pred}}$ can be instantiated as common recommendation objectives, such as binary cross-entropy or pairwise ranking loss.
After local training, we denote the learned item embedding of item $i$ on client $k$ as $\mathbf{e}^{(k)}_{i}$.

\subsubsection{Information Distillation Module}

In \ours, all clients share the same global SIDs and codebooks for feature-indexed communication.
However, each client observes collaborative information from its own local interactions, which can be sparse, biased, and different from those of other clients~\cite{liu2025unbiased}.
If we directly train the codebooks with sparse local interaction loss, such biased local information may be forced into the current SID groups and mixed with unrelated items.
This makes codebook optimization unstable and may harm the alignment required for server aggregation.

To address this issue, \ours introduces an \textbf{information distillation module}.
Instead of directly learning collaborative information on the codebooks, each client first trains item embeddings through local model training.
Then, each client distills the useful collaborative information captured by the learned item embeddings into a local set of feature-indexed codebooks under the globally shared SIDs.
In this way, sparse and biased local information is represented in the same SID space, allowing the server to aggregate the uploaded codebooks.

Specifically, client $k$ maintains a local codebook
\begin{equation}
\mathcal{B}_{k}
=
\Big\{\mathbf{B}_{k}^{(\ell)}\in\mathbb{R}^{M\times d}\Big\}_{\ell=1}^{L}.
\end{equation}
Given the current global SIDs $\mathbf{q}$, the embedding reconstructed from the local codebooks is:
$
\tilde{\mathbf{e}}^{(k)}_{i}
=
\sum_{\ell=1}^{L}
\mathbf{B}_{k}^{(\ell)}
\big[
q_i^{(\ell)}
\big].
$

The locally trained item embedding $\mathbf{e}^{(k)}_i$ acts as the teacher, and the reconstructed embedding $\tilde{\mathbf{e}}^{(k)}_i$ acts as the student.
Client $k$ optimizes its local codebook by minimizing the reconstruction loss:
\begin{equation}
\label{eq:loss_code}
\mathcal{L}^{(k)}_{\mathrm{distill}}(\mathcal{B}_{k})
=
\frac{1}{|\mathcal{I}_k|}
\sum_{i\in\mathcal{I}_k}
\left\|
\mathbf{e}^{(k)}_{i}
-
\tilde{\mathbf{e}}^{(k)}_{i}
\right\|_2^2,
\end{equation}
where $\mathcal{I}_k=\{i\mid (u,i,y)\in\mathcal{D}_k\}$ is the set of items appearing in client $k$'s local dataset.
During this stage, the locally trained item embeddings and the global SID assignment are fixed, and only the local codebook $\mathcal{B}_{k}$ is optimized.
After information distillation, client $k$ uploads $\mathcal{B}_{k}$ to the server for aggregation.

\subsubsection{Server Codebook Aggregation}

After receiving local codebooks from clients, the server aggregates them by weighted averaging.
For each RQ layer $\ell\in\{1,\dots,L\}$, the global codebook is computed as:
\begin{equation}
\label{eq:agg}
\mathbf{B}^{(\ell)}
=
\sum_{k=1}^{K}
\frac{|\mathcal{D}_k|}
{\sum_{j=1}^{K}|\mathcal{D}_j|}
\,
\mathbf{B}_{k}^{(\ell)}.
\end{equation}
The aggregated codebooks $\mathcal{B}$ summarize collaborative information distilled from different clients.
Since $L$, $M$, and $d$ are fixed, the upload size of \ours is bounded and does not grow with the number of items. 

\subsubsection{Local Embedding Update}

After codebook aggregation, the server broadcasts the updated global codebooks $\mathcal{B}$ and SIDs $\mathbf{q}$ to clients.
Each client then refreshes its local item embeddings by reconstructing them from the received codebooks:
\begin{equation}
\label{eq:refresh_item}
\mathbf{e}^{(k)}_i
\leftarrow
\sum_{\ell=1}^{L}
\mathbf{B}^{(\ell)}
\big[
q_i^{(\ell)}
\big].
\end{equation}
The refreshed item embeddings are used as the initialization for the next round of local model training.
This update allows each client to receive item information aggregated from other clients through shared feature-indexed codebooks.

\subsubsection{Self-Refining SID Update Module}

The information distillation module transfers locally learned collaborative information into the shared codebook space, enabling the server to aggregate client information under globally aligned SIDs.
However, this only solves the aggregation problem under the static SID assignment.
Different from conventional RQ-based recommendation, where SIDs are usually generated before training and then kept fixed, the item embeddings in federated recommendation continuously evolve through iterative local training and server aggregation.
As clients update item embeddings from private interactions and the server aggregates distilled codebooks across rounds, the collaborative structure reflected by the global codebooks may gradually change.
Therefore, a static SID assignment may become misaligned with the updated item embeddings and limit the effectiveness of feature-indexed communication.

To address this issue, \ours introduces a \textbf{self-refining SID update module}.
The key idea is to periodically refine SIDs according to the federated item embeddings reconstructed from the aggregated codebooks.

Specifically, the server reconstructs a global embedding for each item $i$ using the current global codebooks and SIDs:
\begin{equation}
\label{eq:global_reconstruct_for_sid}
\bar{\mathbf{e}}_i
=
\sum_{\ell=1}^{L}
\mathbf{B}^{(\ell)}
\big[
q_i^{(\ell)}
\big].
\end{equation}
Here, $\bar{\mathbf{e}}_i$ is not uploaded by any client, but reconstructed on the server from the aggregated codebooks.
It summarizes the collaborative information distilled from different clients under the current SID assignment.

After every $\tau$ communication rounds, the server applies RQ-Kmeans to the reconstructed item embeddings
$\{\bar{\mathbf{e}}_i\}_{i\in\mathcal{I}}$:
\begin{equation}
\label{eq:sid_update}
(\mathbf{q}^{\mathrm{new}}, \mathcal{B}^{\mathrm{new}})
=
\operatorname{RQ\text{-}Kmeans}
\left(
\{\bar{\mathbf{e}}_i\}_{i\in\mathcal{I}}
\right).
\end{equation}
If the current round is not a SID update round, we keep $\mathbf{q}^{\mathrm{new}}=\mathbf{q}$ and $\mathcal{B}^{\mathrm{new}}=\mathcal{B}$.

This module makes SIDs dynamic during federated training rather than static preprocessing tokens.
By periodically refining SIDs based on aggregated codebooks, \ours allows the SIDs and codebooks to gradually adapt to the updated collaborative information without collecting raw interactions or uploading raw local item embeddings.

\subsection{Algorithm}

Our complete training procedure is summarized in Algorithm~\ref{alg:main}.

\begin{algorithm}[t]
\caption{\ours: Feature-Indexed Federated Recommendation}
\label{alg:main}
\small
\begin{algorithmic}[1]
\REQUIRE Clients $\{1,\dots,K\}$ with local datasets $\{\mathcal{D}_k\}$; codebook size $M$; RQ layers $L$; SID update period $\tau$; total rounds $T$.
\STATE Randomly initialize global codebooks $\mathcal{B}^{0}$, and initialize global SIDs $\mathbf{q}^{0}$ randomly or with optional semantic information.
\FOR{round $t=0,1,\dots,T-1$}
    \FOR{each client $k$ in parallel}
        \STATE \textbf{Local training:} update local model parameters $\theta_k$ on $\mathcal{D}_k$ by minimizing Eq.~\eqref{equ:loss_pred}.
        \STATE \textbf{Information distillation:} optimize local codebooks $\mathcal{B}_{k}^{t}$ under global SIDs $\mathbf{q}^{t}$ by minimizing Eq.~\eqref{eq:loss_code}.
        \STATE Upload local codebooks $\mathcal{B}_{k}^{t}$ to the server.
    \ENDFOR
    \STATE \textbf{Codebook aggregation:} aggregate uploaded codebooks to obtain $\mathcal{B}^{t+1}$ by Eq.~\eqref{eq:agg}.
    \IF{$(t+1) \bmod \tau = 0$}
        \STATE Reconstruct global item embeddings from $\mathcal{B}^{t+1}$ and $\mathbf{q}^{t}$ by Eq.~\eqref{eq:global_reconstruct_for_sid}.
        \STATE \textbf{Self-refining SID update:} regenerate SIDs $\mathbf{q}^{t+1}$ and codebooks $\mathcal{B}^{t+1}$ by RQ-Kmeans using Eq.~\eqref{eq:sid_update}.
    \ELSE
        \STATE Keep $\mathbf{q}^{t+1}=\mathbf{q}^{t}$.
    \ENDIF
    \STATE Server broadcasts $\mathcal{B}^{t+1}$ and $\mathbf{q}^{t+1}$ for the next round.
    \STATE Each client reconstructs local item embeddings from $\mathcal{B}^{t+1}$ and $\mathbf{q}^{t+1}$.
\ENDFOR
\end{algorithmic}
\end{algorithm}

\subsection{Privacy Protection}

\ours focuses on changing the communication paradigm of federated recommendation from ID-indexed communication to feature-indexed communication.
Therefore, it is compatible with the privacy protection mechanisms used by existing ID-indexed federated recommendation methods.
Specifically, if a base method applies privacy protection to item embeddings or item updates, \ours can perform codebook distillation on the protected item embeddings and upload the codebooks under the same protection setting.

In this sense, \ours does not require additional access to raw interactions, user embeddings, or unprotected item embeddings.
It only replaces the communicated object from item embeddings to feature-indexed codebooks.
Thus, when the original ID-indexed communication satisfies a given privacy requirement, \ours can inherit the same privacy setting by learning and aggregating codebooks from the protected item information.

\subsection{Communication Cost}

We compare the communication cost of \ours with ID-indexed communication methods such as \textit{FedMF}~\cite{chai2020secure}.
In ID-indexed communication, each client uploads and downloads item embeddings organized by item IDs.
Since the item embedding table has size $n_i \times d$, where $n_i$ is the number of items and $d$ is the embedding dimension, the communication cost scales as $\mathcal{O}(n_i d)$.

In contrast, \ours communicates feature-indexed codebooks instead of full item embedding tables.
With $L$ RQ layers and $M$ code embeddings in each codebook, the codebook communication cost scales as $\mathcal{O}(L M d)$.
When a SID update is performed, the server additionally broadcasts the updated SID assignment, whose cost is $\mathcal{O}(n_i L)$.
The download cost of \ours becomes $\mathcal{O}\left(LMd + n_i L\right)$.
Because $L$ and $M$ are fixed hyperparameters and $L \ll d$, the SID table is much cheaper than transmitting full item embeddings.
In our experiments, we set $LM \ll n_i$, so the total download cost $\mathcal{O}(LMd+n_iL)$ is consistently smaller than that of the ID-indexed item embedding table.
Table~\ref{tab:comres} provides a comparison of the download cost.

\section{Theoretical Analysis}
\label{sec:theory}

We provide a concise analysis to explain why feature-indexed communication can reduce the influence of noisy updates.
This proof leads to an intuitive conclusion: independent item embeddings are less resistant to noise than feature embeddings shared by multiple items.
This is because ID-indexed aggregation updates each item embedding only with independent item signals, while feature-indexed aggregation pools signals from all items sharing the same feature embedding.
We show that this feature-indexed aggregation can strictly reduce the variance of noisy updates.

\subsection{Noise Reduction by Feature-Indexed Aggregation}

Consider a set of items $\mathcal{I}$.
In ID-indexed communication, each item $i$ is treated as an independent communication unit.
Let $\mathcal{K}(i)$ denote the set of clients that contribute updates for item $i$, and let $n_i=|\mathcal{K}(i)|$.
Assume that client $k$ observes a noisy item update:
\begin{equation}
\mathbf{z}_{i}^{(k)}
=
\mathbf{z}_{i}^{\star}
+
\boldsymbol{\epsilon}_{i}^{(k)},
\qquad
\mathbb{E}[\boldsymbol{\epsilon}_{i}^{(k)}]=\mathbf{0},
\qquad
\mathbb{E}\|\boldsymbol{\epsilon}_{i}^{(k)}\|_2^2=\sigma^2,
\end{equation}
where $\mathbf{z}_{i}^{\star}$ is the ideal item embedding without noise information, and $\boldsymbol{\epsilon}_{i}^{(k)}$ denotes noise caused by sparse interactions, stochastic optimization, or noisy feedback.

The ID-indexed estimator for item $i$ is:
\begin{equation}
\widehat{\mathbf{z}}_{i}^{\mathrm{ID}}
=
\frac{1}{n_i}
\sum_{k\in\mathcal{K}(i)}
\mathbf{z}_{i}^{(k)}.
\end{equation}

In feature-indexed communication, items are mapped to shared feature indices.
Let $q_i\in\{1,\dots,M\}$ denote the feature index of item $i$, and let $\mathcal{I}_m=\{i\in\mathcal{I}\mid q_i=m\}$ be the set of items sharing feature index $m$. Thus, items in $\mathcal{I}_m$ share the same underlying feature embedding $\mathbf{b}_m^{\star}$.
For each item $i\in\mathcal{I}_m$, its client update can be written as:
\begin{equation}
\mathbf{z}_{i}^{(k)}
=
\mathbf{b}_{m}^{\star}
+
\boldsymbol{\epsilon}_{i}^{(k)}.
\end{equation}

The feature-indexed estimator aggregates all updates assigned to feature index $m$:
\begin{equation}
\widehat{\mathbf{b}}_{m}^{\mathrm{F}}
=
\frac{1}{N_m}
\sum_{i\in\mathcal{I}_m}
\sum_{k\in\mathcal{K}(i)}
\mathbf{z}_{i}^{(k)},
\qquad
N_m=\sum_{i\in\mathcal{I}_m} n_i.
\end{equation}

\begin{theorem}
\label{thm:feature_noise}
Assume that the noise terms are independent and zero-mean with variance $\sigma^2$.
For an item $i$ with feature index $q_i=m$, let 
$\mathcal{E}_{\mathrm{ID}}(i)$ denote the expected aggregation error under ID-indexed aggregation, and let 
$\mathcal{E}_{\mathrm{F}}(m)$ denote the expected aggregation error under feature-indexed aggregation.
Then,
\begin{equation}
\label{eq:noise_compare}
\mathcal{E}_{\mathrm{F}}(m)
\le
\mathcal{E}_{\mathrm{ID}}(i).
\end{equation}
The inequality is strict when the feature index $m$ is shared by at least one additional item with client updates.
\end{theorem}

\begin{proof}
According to the ID-indexed estimator, its expected aggregation error is:
\begin{equation}
\begin{aligned}
\mathcal{E}_{\mathrm{ID}}(i)
&=
\mathbb{E}
\left\|
\widehat{\mathbf{z}}_{i}^{\mathrm{ID}}
-
\mathbf{z}_{i}^{\star}
\right\|_2^2  \\
&=
\mathbb{E}
\left\|
\frac{1}{n_i}
\sum_{k\in\mathcal{K}(i)}
\boldsymbol{\epsilon}_{i}^{(k)}
\right\|_2^2 .
\end{aligned}
\end{equation}
Since the noise terms are independent and zero-mean, we have:
\begin{equation}
\mathcal{E}_{\mathrm{ID}}(i)
=
\frac{1}{n_i^2}
\sum_{k\in\mathcal{K}(i)}
\mathbb{E}
\left\|
\boldsymbol{\epsilon}_{i}^{(k)}
\right\|_2^2
=
\frac{\sigma^2}{n_i}.
\end{equation}

Similarly, according to the feature-indexed estimator, its expected aggregation error is:
\begin{equation}
\begin{aligned}
\mathcal{E}_{\mathrm{F}}(m)
&=
\mathbb{E}
\left\|
\widehat{\mathbf{b}}_{m}^{\mathrm{F}}
-
\mathbf{b}_{m}^{\star}
\right\|_2^2 \\
&=
\mathbb{E}
\left\|
\frac{1}{N_m}
\sum_{j\in\mathcal{I}_m}
\sum_{k\in\mathcal{K}(j)}
\boldsymbol{\epsilon}_{j}^{(k)}
\right\|_2^2 .
\end{aligned}
\end{equation}
By independence and zero mean, we obtain:
\begin{equation}
\mathcal{E}_{\mathrm{F}}(m)
=
\frac{1}{N_m^2}
\sum_{j\in\mathcal{I}_m}
\sum_{k\in\mathcal{K}(j)}
\mathbb{E}
\left\|
\boldsymbol{\epsilon}_{j}^{(k)}
\right\|_2^2
=
\frac{\sigma^2}{N_m}.
\end{equation}

Since item $i$ belongs to $\mathcal{I}_m$, the feature index $m$ aggregates at least the updates of item $i$:
\begin{equation}
N_m
=
\sum_{j\in\mathcal{I}_m} n_j
\ge
n_i.
\end{equation}
Therefore,
\begin{equation}
\mathcal{E}_{\mathrm{F}}(m)
=
\frac{\sigma^2}{N_m}
\le
\frac{\sigma^2}{n_i}
=
\mathcal{E}_{\mathrm{ID}}(i).
\end{equation}
The equality holds when $N_m=n_i$.
When feature index $m$ is shared by at least one additional item with client updates, we have $N_m>n_i$, and thus:
\begin{equation}
\mathcal{E}_{\mathrm{F}}(m)
<
\mathcal{E}_{\mathrm{ID}}(i).
\end{equation}
\end{proof}

Theorem~\ref{thm:feature_noise} shows a core advantage of feature-indexed communication.
ID-indexed aggregation averages noisy updates only within the independent item ID, while feature-indexed aggregation pools updates from all items sharing the same feature index.
Therefore, the effective number of observations increases from $n_i$ to $N_m$, reducing the noise variance from $\sigma^2/n_i$ to $\sigma^2/N_m$.
This explains why feature-indexed communication can improve robustness to noisy and sparse ID-indexed feedback.

\section{Experiments}

In this section, we introduce and analyze the following research questions (RQs):
\begin{itemize}
    \item \textbf{RQ1:} Does \ours outperform representative federated recommendation baselines in recommendation accuracy?

    \item \textbf{RQ2:} Do our information distillation module and self-refining SID update module contribute to the performance of \ours?

    \item \textbf{RQ3:} Can feature-indexed communication address the efficiency and effectiveness limitations of ID-indexed communication?

    \item \textbf{RQ4:} Is the feature-indexed codebook construction in \ours more effective than directly using semantic information or alternative codebook construction methods?
\end{itemize}

\subsection{Experimental Settings}

\begin{table}[t]
\centering
\caption{Statistics of our datasets.
}
\resizebox{0.8\columnwidth}{!}
{
\begin{tabular}{l|ccc}
\toprule
datasets & \# Users & \# Items & \# Interactions \\
\midrule
ML-100K  & 943      & 1682     & 100000          \\
ML-1M    & 6040     & 3706     & 1000209         \\
Steam    &23310     &5237      &525922             \\
Movies    &16641    &20000      &130342             \\
VideoGames    &13107    &20000      &112043         \\
\bottomrule
\end{tabular}
}
\label{tab:dataset}
\end{table}

\begin{table*}[t]
\centering
\small
\setlength{\tabcolsep}{4pt}
\caption{Overall comparison on five datasets. The boldface indicates the best result and the underline indicates the second-best.}
\label{tab:result_table}
\resizebox{\textwidth}{!}{
\begin{tabular}{l l c ccccccc cc}
\toprule
\multirow{2}{*}{Dataset} & \multirow{2}{*}{Metric}
& \multicolumn{1}{c}{No Agg.}
& \multicolumn{7}{c}{ID-Indexed Aggregation}
& \multicolumn{2}{c}{Ours} \\
\cmidrule(lr){3-3}\cmidrule(lr){4-10}\cmidrule(lr){11-12}
& & MF
& FedMF & FedNCF & PFedRec & GPFedRec & FedRAP & FedCIA & FedCA
& RQFedRec-R & RQFedRec-S \\
\midrule

\multirow{3}{*}{ML-100K}
& Recall@10 & 0.0756 & 0.1813 & 0.1680 & 0.1681 & 0.1162 & 0.0600 & 0.1903 & 0.1184 & \underline{0.1928} & \textbf{0.1973} \\
& MRR@10    & 0.3232 & 0.5501 & 0.5568 & 0.5685 & 0.4550 & 0.0600 & 0.5855 & 0.4107 & \underline{0.5973} & \textbf{0.6168} \\
& NDCG@10   & 0.1504 & 0.3198 & 0.3170 & 0.3233 & 0.2395 & 0.1260 & 0.3440 & 0.2008 & \underline{0.3533} & \textbf{0.3650} \\
\midrule

\multirow{3}{*}{ML-1M}
& Recall@10 & 0.0641 & 0.1285 & 0.1298 & 0.1177 & 0.0798 & 0.0592 & 0.1392 & 0.0806 & \underline{0.1497} & \textbf{0.1517} \\
& MRR@10    & 0.3601 & 0.5664 & 0.5590 & 0.5367 & 0.4269 & 0.3082 & 0.5947 & 0.4167 & \underline{0.6176} & \textbf{0.6300} \\
& NDCG@10   & 0.1816 & 0.3254 & 0.3265 & 0.3192 & 0.2396 & 0.1572 & 0.3611 & 0.1989 & \underline{0.3750} & \textbf{0.3813} \\
\midrule

\multirow{3}{*}{Steam}
& Recall@10 & 0.0652 & 0.0611 & 0.0535 & 0.0492 & 0.0562 & 0.0512 & 0.0752 & 0.0538 & \underline{0.0779} & \textbf{0.0809} \\
& MRR@10    & \textbf{0.1317} & 0.0913 & 0.0737 & 0.0799 & 0.0906 & 0.1095 & 0.1143 & 0.1068 & 0.1157 & \underline{0.1189} \\
& NDCG@10   & 0.0649 & 0.0508 & 0.0429 & 0.0429 & 0.0489 & 0.0522 & 0.0644 & 0.0528 & \underline{0.0653} & \textbf{0.0673} \\
\midrule

\multirow{3}{*}{Movies}
& Recall@10 & 0.1110 & 0.1194 & 0.1216 & 0.0858 & 0.1158 & 0.0531 & 0.1146 & 0.1018 & \textbf{0.1460} & \underline{0.1349} \\
& MRR@10    & 0.0996 & 0.0903 & 0.0914 & 0.0571 & 0.0881 & 0.0576 & 0.0894 & 0.0812 & \textbf{0.1117} & \underline{0.1057} \\
& NDCG@10   & 0.0804 & 0.0781 & 0.0799 & 0.0521 & 0.0760 & 0.0426 & 0.0768 & 0.0686 & \textbf{0.0978} & \underline{0.0913} \\
\midrule

\multirow{3}{*}{VideoGames}
& Recall@10 & 0.0163 & \underline{0.0414} & 0.0198 & 0.0128 & 0.0312 & 0.0018 & 0.0300 & 0.0193 & \textbf{0.0483} & 0.0405 \\
& MRR@10    & 0.0200 & \underline{0.0323} & 0.0156 & 0.0097 & 0.0282 & 0.0019 & 0.0227 & 0.0199 & \textbf{0.0395} & 0.0290 \\
& NDCG@10   & 0.0139 & \underline{0.0270} & 0.0127 & 0.0081 & 0.0219 & 0.0014 & 0.0190 & 0.0147 & \textbf{0.0323} & 0.0259 \\
\bottomrule
\end{tabular}
}
\end{table*}

\subsubsection{Datasets}

Our experiments are conducted on five widely used datasets: ML-100K~\cite{harper2015movielens}, ML-1M~\cite{harper2015movielens}, Steam~\cite{Steam}, Movies~\cite{mcauley2015image} and VideoGames~\cite{mcauley2015image}.
These datasets are split into training and test sets in an 8:2 ratio, with 10\% of the training set used as a validation set.
Following a general federated recommendation setting~\cite{han2025fedcia}, we do not restrict each client to a single user. A client can represent a user device or a user group. This setting includes the single user case as a special case and better matches our focus on communication under large item catalogs. Therefore, we partition each dataset into 100 clients and evaluate all methods under the same protocol.
The statistics of these datasets are shown in Table~\ref{tab:dataset}.

\subsubsection{Compared Methods}

We compare two popular federated recommendation algorithms, FedMF~\cite{chai2020secure} and FedNCF~\cite{perifanis2022federated}, and five methods that focus on user personalization, PFedRec~\cite{zhang2023dual}, GPFedRec~\cite{zhang2024gpfedrec}, FedRAP~\cite{li2024federated}, FedCIA~\cite{han2025fedcia} and FedCA~\cite{zhang2026beyond}.

Our primary goal is to improve model performance within the same communication budget. Therefore, we mainly compare state-of-the-art (SOTA) methods rather than methods specifically designed for communication efficiency.

\subsubsection{Evaluation Metrics}

We use Recall~\cite{herlocker2004evaluating}, Mean Reciprocal Rank (MRR)~\cite{voorhees1999trec} and Normalized Discounted Cumulative Gain (NDCG)~\cite{jarvelin2002cumulated} as the primary evaluation metrics. These are popular metrics in the federated recommendation scenario. We set $K=10$ for these metrics.

\subsubsection{Implementation Details}

Our method supports two SID initialization strategies.
The first is \textbf{random initialization}, denoted as \textbf{RQFedRec-R}, where SIDs are randomly initialized without using any external semantic information.
The second is \textbf{semantic initialization}, denoted as \textbf{RQFedRec-S}, where public semantic information is used to initialize SIDs.

For semantic initialization, following existing large language model–based recommendation work~\cite{ren2024representation,liu2025improving}, we obtain a semantic embedding for each item by first generating a structured item feature profile with the OpenAI gpt-3.5-turbo model using item titles and descriptions as input, and then encoding the resulting profile with text-embedding-ada002. We also provide the encoded semantic embeddings in our reproduction code.
We apply RQ-Kmeans to these semantic embeddings to initialize SIDs. We set different codebook sizes according to the dataset scale, while ensuring that the total number of embeddings across all codebooks is smaller than the number of items in the dataset.

It is worth noting that many personalized federated learning works construct the recall set for validation and testing with one positive item and 99 negative items. This setting can lead to unusually high performance in some cases. In our experiments, we follow the standard recommendation setup~\cite{he2020lightgcn} and evaluate against all items as the recall set. Therefore, our results are not directly comparable to those works.

We implement \ours using PyTorch on an NVIDIA Tesla T4 GPU. We use MF as our backbone model. For fair comparison, we follow the same parameter ranges as \ours and the corresponding baselines. For hyperparameter tuning, we search learning rates in $\{0.1, 0.01, 0.001\}$, weight decay in $\{0, 10^{-6}, 10^{-3}\}$, batch sizes in $\{256, 1024, 4096\}$, and SID update periods $\tau$ in $\{5, 10, 20\}$. We use Adam for local training, and detailed settings are provided in our reproduction code.

\subsection{Method Comparison (RQ1)}

We compare \ours with all federated baselines, as shown in Table~\ref{tab:result_table}. We also include methods without federated aggregation, and draw the following observations.

1) Federated learning methods outperform those without aggregation, showing that sharing information across clients effectively improves recommendation performance.
2) Some personalized baselines are optimized for single user clients, so their user-level personalization advantages may weaken under our general setting.
In contrast, RQFedRec focuses on improving communication and thus remains effective in this broader setting.
3) \ours-R achieves strong performance compared with existing federated recommendation baselines. This shows that the improvement of \ours does not come from semantic priors, but mainly benefits from feature-indexed communication, information distillation, and self-refining SID update.
4) \ours-S outperforms \ours-R and other federated recommendation baselines on ML-100K, ML-1M, and Steam, showing that \ours can effectively leverage reliable semantic information.
However, when semantic information is noisy or weakly aligned with user behavior, \ours-R can perform better, as observed on Movies and VideoGames.

In summary, our method generally outperforms existing baselines without any additional semantic information, while providing a reasonable scheme for utilizing additional semantic information.

\subsection{Ablation Study (RQ2)}
\label{sec:ablation}

\begin{figure}[t]
\centering
\includegraphics[width=1\columnwidth]{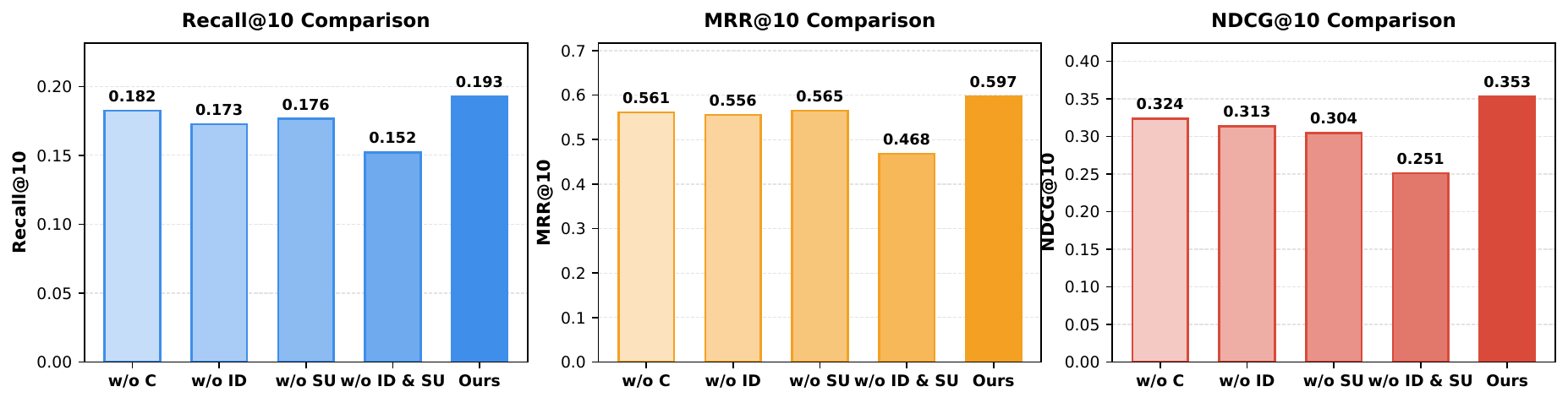}
\caption{The ablation study of \ours on the ML-100K dataset.}
\label{fig:abla}
\end{figure}

We conduct ablation studies to evaluate the contribution of the key designs in \ours.
Specifically, we compare \ours with the following variants:
1) \textbf{w/o C}: removing feature-indexed codebook communication and using the original ID-indexed item embedding communication.
2) \textbf{w/o ID (Information Distillation)}: removing the information distillation module and directly optimizing codebooks with the recommendation loss.
3) \textbf{w/o SU (SID Update)}: removing the self-refining SID update module and keeping the initial SIDs fixed during training.
4) \textbf{w/o ID \& SU}: removing both information distillation and self-refining SID update.
5) \textbf{\ours}: the full model with both modules.
We conduct experiments on ML-100K, and the results are shown in Figure~\ref{fig:abla}.

From the results, we observe that the full \ours consistently outperforms all ablated variants.
1) \textbf{w/o C} performs worse than \ours on all metrics, confirming that feature-indexed codebook communication is more effective than directly communicating ID-indexed item embeddings.
2) removing either information distillation or self-refining SID update leads to clear performance degradation.
The performance drop of \textbf{w/o ID} indicates that directly optimizing codebooks from sparse local interactions is insufficient, and locally learned item embeddings provide useful guidance for transferring collaborative information into codebooks.
The performance drop of \textbf{w/o SU} shows that fixed SIDs cannot fully match the collaborative information that evolves during federated training.
Finally, \textbf{w/o ID \& SU} further degrades performance, suggesting that the two modules are complementary.
These results verify the necessity of feature-indexed communication and demonstrate that both proposed modules are important for constructing effective and adaptive communication in \ours.

\subsection{Analysis of Feature-Indexed Communication (RQ3)}

In this section, we mainly analyze whether our method addresses the problems we identified.

\subsubsection{Communication Cost}

\begin{table}[t]
\centering
\caption{Comparison of download communication cost across different methods. Percentage denotes the communication cost relative to the baseline.}
\resizebox{1\columnwidth}{!}
{
\begin{tabular}{l|l|ccc}
\toprule
Dataset & Method & Parameter & Cost & Percentage  \\ \hline
\multirow{2}{*}{ML-100K}& FedMF & $d=128, n_i = 1682$  & $215296$ & 100\%\\
&\ours  & $d=128, M = 512, L=3$ & $\textbf{201654}$ &\textbf{ 93\%} \\ \hline
\multirow{2}{*}{ML-1M}&FedMF  & $d=128, n_i = 3706$  & $474368$ & 100\%\\
&\ours  & $d=128, M = 1024, L=3$ & $\textbf{404334}$ & \textbf{85\%} \\ \hline
\multirow{2}{*}{Steam}&FedMF  & $d=128, n_i = 5237$  & $670336$ & 100\%\\
&\ours  & $d=128, M = 1024, L=3$ & $\textbf{408927}$ & \textbf{61\%} \\ \hline
\multirow{2}{*}{Movies} & FedMF  & $d=128, n_i = 20000$  & $2560000$ & 100\%\\
&\ours  & $d=128, M = 4096, L=3$ & $\textbf{1632864}$ & \textbf{64\%} \\ \hline
\multirow{2}{*}{VideoGames} & FedMF & $d=128, n_i = 20000$  & $2560000$ & 100\%\\
&\ours  & $d=128, M = 4096, L=3$ & $\textbf{1632864}$ &\textbf{64\%} \\ 
\bottomrule
\end{tabular}
}
\label{tab:comres}
\end{table}

This section analyzes the communication cost of our method, including the theoretical communication cost under different paradigms, and reports the specific settings and communication cost of our method.
Since downloading incurs the larger communication volume in \ours, we report its download cost in Table~\ref{tab:comres}.
\ours reduces communication on all datasets while achieving the performance reported in Table~\ref{tab:result_table}; further compression is possible when efficiency is prioritized over accuracy.

\subsubsection{Collaborative Information Communication}

\begin{figure}[t]
\centering
\includegraphics[width=0.9\columnwidth]{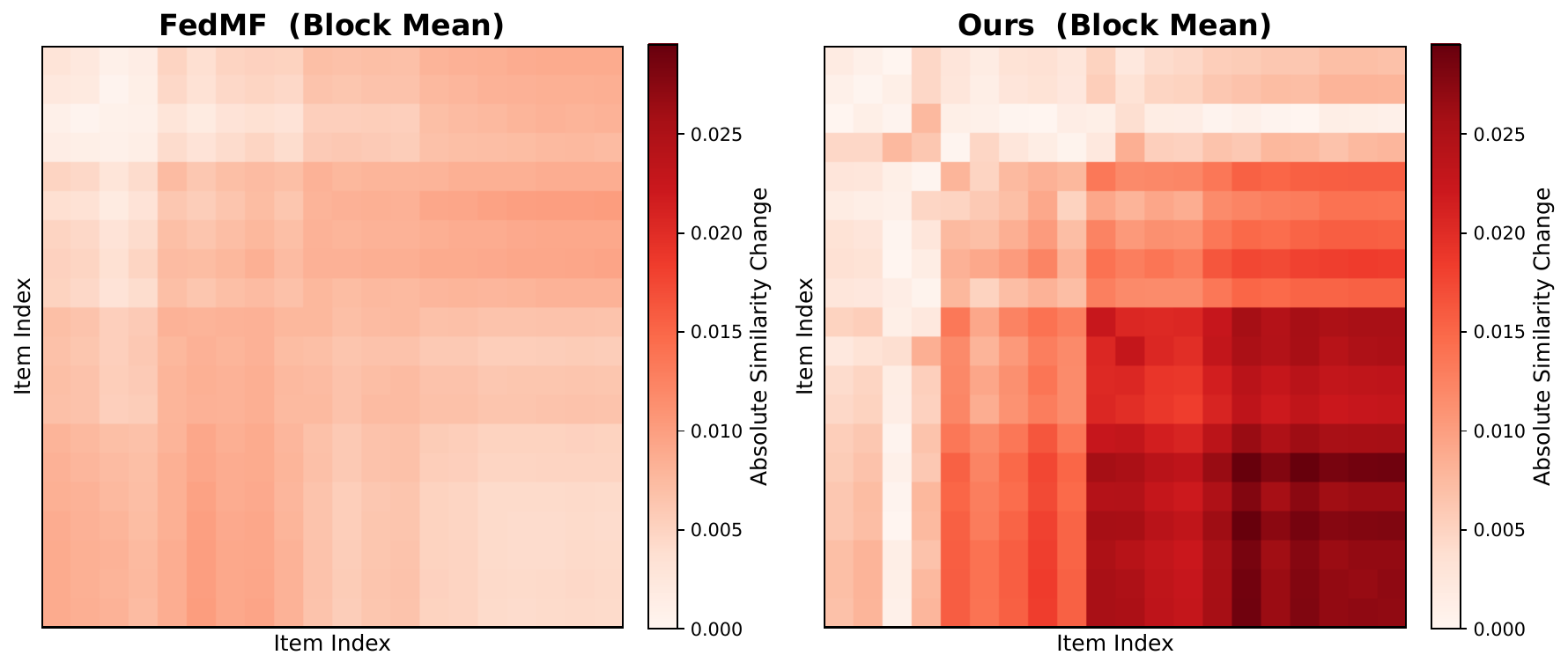}
\caption{Visualization of item-item collaborative information updates after communication. Darker colors indicate larger information.}\label{fig:com}
\end{figure}

We further analyze whether feature-indexed communication can transmit stronger collaborative information than ID-indexed communication.
We compare \ours with FedMF on ML-100K by averaging the absolute changes in item-item cosine similarity over 10 communication rounds.
For visualization, we divide items into 20 consecutive index blocks; each heatmap cell shows the average similarity change between two blocks, with both methods using the same color scale.

Figure~\ref{fig:com} shows darker regions for \ours in most item-block pairs, indicating that shared codebooks transmit stronger collaborative information and enable larger changes in item relationships across rounds.

\subsubsection{Sensitivity to Noise}

\begin{table}[t]
\centering
\caption{Comparison of different noise ratios on the ML-100K dataset.}
\resizebox{1\columnwidth}{!}
{
\begin{tabular}{l|l|ccc}
\toprule
Noise ratio & Method & Recall@10 & MRR@10 & NDCG@10 \\ \hline
\multirow{2}{*}{0.05}
& FedMF & 0.1641 & 0.5302 & 0.2951 \\
& \ours & \textbf{0.1832} & \textbf{0.5635} & \textbf{0.3330} \\ \hline
\multirow{2}{*}{0.10}
& FedMF & 0.1561 & 0.4959 & 0.2747 \\
& \ours & \textbf{0.1715} & \textbf{0.5338} & \textbf{0.3073} \\ \hline
\multirow{2}{*}{0.15}
& FedMF & 0.1341 & 0.4588 & 0.2444 \\
& \ours & \textbf{0.1495} & \textbf{0.4855} & \textbf{0.2712} \\
\bottomrule
\end{tabular}
}
\label{tab:noise}
\end{table}

In this section, we analyze the sensitivity of our method to noise. Specifically, we simulate user misclicks by injecting noisy interactions into the raw data with different noise ratios, and then train the models on these noisy clicks. We compare our method with the traditional federated learning baseline FedMF on ML-100K, and the results are reported in Table~\ref{tab:noise}.

As shown in the table, our method achieves better performance under all settings, indicating strong robustness. Moreover, as the noise ratio increases, the performance drop of our method is less pronounced, further demonstrating its resistance to noisy data.

\subsubsection{Runtime Analysis}

\begin{table}[t]
\centering
\caption{Runtime comparison on ML-100K.}
\resizebox{1\columnwidth}{!}{
\begin{tabular}{l|ccccc|c}
\toprule
Method & Avg./Round & Train & Upload & Refresh & Download & vs. FedMF \\
\midrule
FedMF     & 7.09s  & 6.93s  & 0.14s & 0.00s & 0.01s  & 1.00$\times$ \\
\ours     & 15.81s & 14.29s & 0.23s & 1.00s & 0.29s  & 2.23$\times$ \\
FedCIA    & 20.09s & 6.82s  & 0.19s & 0.00s & 13.08s & 2.83$\times$ \\
\bottomrule
\end{tabular}
}
\label{tab:runtime}
\end{table}

We further report the runtime cost of different methods in Table~\ref{tab:runtime}.
Compared with FedMF, \ours introduces additional computation for codebook learning and SID/codebook refresh.
Therefore, the average time per communication round increases.
However, the runtime cost of \ours remains practical.
Although \ours is slower than the most basic FedMF baseline, it is still faster than some recent federated recommendation methods such as FedCIA.
This shows that the additional cost of \ours is moderate and acceptable, especially considering its improved recommendation performance and reduced communication cost.

\subsection{Analysis of Feature-Indexed Codebook Construction (RQ4)}
\label{sec:rq4}

We examine whether the improvement of \ours merely comes from external semantics or a compact codebook by comparing it with LLM-based federated methods and alternative codebook construction methods.

\subsubsection{Semantic Information Utilization Analysis}

\begin{table}[t]
\centering
\caption{Comparison with LLM-based federated recommendation methods on ML-100K.}
\resizebox{0.95\columnwidth}{!}
{
\begin{tabular}{l|ccc}
\toprule
 Method     & Recall@10           & MRR@10          & NDCG@10          \\ 
\midrule
FedRLM\_con     & 0.1772          & 0.5775          & 0.3364           \\
FedRLM\_gen     & 0.1890          & 0.5622          & 0.3308           \\ 
FedAlphaFuse    & 0.1836          & 0.5840          & 0.3484           \\ 
\midrule
\ours & \textbf{0.1973} & \textbf{0.6168} & \textbf{0.3650} \\
\bottomrule
\end{tabular}
}
\label{tab:LLM}
\end{table}

Since \ours can optionally use public semantic information for SID initialization, we examine whether its improvement simply comes from LLM-based item information.
Table~\ref{tab:LLM} compares \ours with federated adaptations of RLM\_con and RLM\_gen~\cite{ren2024representation}, and AlphaFuse~\cite{hu2025alphafuse}.
LLM-based semantic representations may be entangled~\cite{zhoudisentangling} or misaligned with collaborative signals, and therefore may not consistently improve federated recommendation.
In contrast, \ours uses semantics only for SID initialization and subsequently refines its codebooks and SIDs, showing that its gains primarily arise from dynamic feature-indexed communication.

\subsubsection{Codebook Construction Analysis}

\begin{table}[t]
\centering
\caption{Comparison with different codebook construction methods on ML-100K.}
\resizebox{0.95\columnwidth}{!}
{
\begin{tabular}{l|ccc}
\toprule
 Method     & Recall@10           & MRR@10          & NDCG@10          \\ 
\midrule
Hash-based codebook & 0.0324          & 0.1250          & 0.0475       \\
PQ-based codebook   & 0.1811          & 0.5735          & 0.3239       \\
RQ-VAE codebook     & 0.1901          & 0.6085          & 0.3554       \\ 
\midrule
\ours & \textbf{0.1973} & \textbf{0.6168} & \textbf{0.3650} \\
\bottomrule
\end{tabular}
}
\label{tab:codebook}
\end{table}

We further compare \ours with other codebooks, including hash-based, PQ-based, and RQ-VAE codebooks. The results are reported in Table~\ref{tab:codebook}.
Hashing ignores relationships among item embeddings, while PQ and RQ-VAE cannot adapt to collaborative structures that evolve during federated training.
In contrast, \ours distills codebooks and dynamically refines SIDs, demonstrating that effective feature-indexed communication requires both compact representation and adaptation to evolving collaborative signals.

\section{Conclusion}
\label{sec:conclusion}

In this paper, we proposed \ours, an RQ-based federated recommendation framework that shifts conventional item embedding communication from an ID-indexed paradigm to a feature-indexed codebook paradigm.
To adapt feature-indexed communication to federated recommendation, \ours introduces information distillation to enable globally aligned aggregation and a self-refining SID update to capture evolving collaborative information.
Extensive experiments show that \ours achieves higher recommendation accuracy and lower communication cost without relying on semantic priors, while still benefiting from public semantics when available.

\begin{acks}

This work is supported by National Natural Science Foundation of China (NSFC) under the Grant No. 62372113, and Major Project of the National Social Science Fund of China (NSSFC) under the Grant No. 25\&ZD260. Peng Zhang \& Tun Lu are with the College of Computer Science and Artificial Intelligence, Fudan University. Tun Lu is also affiliated to the Silver-X MOE Philosophy \& Social Sciences Laboratory, Fudan Institute on Aging, and Shanghai Key Laboratory of Data Science, Fudan University.

\end{acks}

\clearpage
\section*{GenAI Usage Disclosure}

This work uses OpenAI gpt-3.5-turbo to generate structured item feature profiles from public item titles and descriptions, and text-embedding-ada-002 to encode these profiles for semantic SID initialization, as described in the experimental settings. Generative AI tools were also used for language polishing of author-written text. They were not used to generate experimental results or scientific conclusions. All method design, experiments, analyses, and final content were conducted, reviewed, and verified by the authors.

\bibliographystyle{ACM-Reference-Format}
\balance
\bibliography{ref}

\end{document}